\documentclass[a4paper,fleqn,usenatbib]{mnras}

\usepackage{mathptmx}
\usepackage{txfonts}
\usepackage{bold-extra}
\usepackage[T1]{fontenc}
\usepackage[T1]{fontenc}
\usepackage{ae,aecompl}

\usepackage{graphicx}	

\newcommand{\TS}{3FGL~J0909\ }
\newcommand{\ts}{3FGL~J0909}   
\newcommand{\kms}{km~s$^{-1}$}
\newcommand{\ergscmA}{erg~s$^{-1}$cm$^{-2}$\AA$^{-1}$}

\newcommand{\ergscm}{erg~s$^{-1}$cm$^{-2}$}

\title[Redshift of 3FGL J0909.0+2310]{On the Redshifts of the
BL Lac 3FGL~J0909.0+2310 and its Close Companion}

\author[D. Rosa-Gonz\'alez  et al.]{
D. Rosa-Gonz\'alez,$^{1}$\thanks{E-mail: danrosa@inaoep.mx}
S. Couti\~no de Le\'on,$^{1}$, Y. D. Mayya,$^{1}$ A. Carrami\~nana,$^{1}$ \newauthor
I. Aretxaga,$^{1}$ J. Becerra Gonz\'alez,$^{2,3}$ A. Furniss,$^{4}$ E. Terlevich,$^{1}$  O. Vega,$^{1}$\newauthor
J. M\'endez-Abreu,$^{5}$ J. Le\'on Tavares,$^{1}$ A. L. Longinotti,$^{1}$ and R. Terlevich$^{1}$ 
\\
$^{1}$Instituto Nacional de Astrof\'\i sica, Optica y Electr\'onica,
Tonantzintla 72840, Puebla, Mexico \\
$^{2}$NASA Goddard Space Flight Center, Greenbelt, MD 20771, USA\\
$^{3}$Department of Physics and Department of Astronomy, University of Maryland,
College Park, MD 20742, USA \\
$^{4}$
Physics Department, California State University East Bay, Hayward, CA 94542, USA
\\
$^{5}$School of Physics and Astronomy, University of St Andrews, North Haugh, St  Andrews, KY169SS, UK \\
}
\date{Accepted XXX. Received YYY; in original form ZZZ}

\pubyear{2016}

\begin{document}
\label{firstpage}
\pagerange{\pageref{firstpage}--\pageref{lastpage}}
\maketitle

\begin{abstract}
We report  on the redshift of the BL Lac object  3FGL~J0909.0+2310
based on observations obtained with the OSIRIS Multi Object Spectrograph (MOS)
mounted on the 10.4-m Gran Telescopio Canarias.
A redshift of 0.432$\pm$0.002 was obtained by the identification
of three absorption features (Ca \textsc{II} K\&H and G-band) detected in 
the spectrum of the BL Lac host galaxy. 
The closest object to the BL Lac at an angular separation of 
3.8\arcsec ($\sim$21 kpc at this distance) has a similar redshift of 0.431$\pm$0.002.
This companion galaxy could be the most likely cause of the nuclear activity 
as postulated by studies based on more extended data sets
and cosmological models.
MOS allows us to study the object's neighbourhood  
within a field of view of approximately 7'$\times$2' and we find two small groups
of galaxies at redshifts 0.28 and 0.39 which are probably not related to the 
activity of 3FGL~J0909.0+2310. 
\end{abstract}




\section{Introduction}

Blazars, in particular BL Lac objects~\citep[see][for a recent review]{2014Falomo}, 
are  described as extreme active galactic nuclei (AGN) in which the relativistic jets
originating close to the central massive black hole (BH) are pointing directly to 
the observer~\citep{1978Blandford}. 
The highly beamed energy released by the central BH 
across the entire electromagnetic
spectrum makes them visible at large distances, 
but at the same time dilutes severely any spectral feature from its host
galaxy~\citep{2011Leon,2013Furniss}.
\citet{2013Shaw}  recently compiled 
the largest sample of $\gamma$-ray selected BL Lac objects (BLLs) with spectroscopic redshifts.
They  used literature data and their own observations in 4 and 10 meter class telescopes
resulting in successful redshift measurements for only 44\% of the 475 studied BLLs. 

3FGL~J0909.0+2310, also named  SDSS J090900.62+231112.9, (\TS hereafter)
was first detected by the NRAO Green Bank telescope~\citep{1991Becker}
and classified as a radio-loud AGN by~\citet{1997Brinkmann}.
The classification as a BL Lac comes from 
high angular resolution radio images obtained with the VLA 
of targets with detection in both 
the NRAO Green Bank telescope and the ROSAT All-Sky Survey~\citep{1997Laurent}.
First attempts to obtain the redshift of \TS are from  
\cite{2011Allen} who, applying a non-negative matrix factorisation method 
to the SDSS spectrum, estimated $z=1.1844$.
\citet{2012Aliu} and \citet{2013Shaw} proposed a lower limit of $z>0.43$
based on the
detection of  the 2800~\AA\, MgII doublet
in the SDSS spectrum 
which could be either intrinsic, or due to the presence of absorbers in the line of sight.

\TS appears in gamma rays in the 
First Catalog of point sources detected 
by the Fermi Large Area Telescope~\citep[1FGL,][]{2010Abdo} with a flux in the 100~MeV to 100~GeV 
range of $1.20\pm0.38\times10^{-11}\,$\ergscm.
It is also listed in the second~\citep[2FGL,][]{2012Nolan} and third~\citep[3FGL,][]{2015Acero}
Fermi releases 
with fluxes of $1.14\pm0.24\times10^{-11}$ and
$7.59\pm1.29\times10^{-12}$\,\ergscm\, respectively.
These observations report a hard power law ($\frac{dN}{dE}\propto E^{-\Gamma}$)
spectral index of $\Gamma\sim 1.7$. 
The object is included in the 
First Fermi-LAT Catalog of Sources above 10 GeV~\cite[1FHL,][]{2013Ackermann},
with a flux of $1.00\pm0.50\times10^{-11}$\,\ergscm\, and 
a spectral index in the 10-500 GeV energy range of $\Gamma=1.90\pm0.36$. 
Sources with hard spectral index and 
detected by Fermi above 10 GeV 
are targets of interest for follow up studies at very high
energies (VHE; E $> 100$ GeV).
In particular \TS has been observed by 
the imaging Cherenkov telescopes
VERITAS but only upper limits to the
fluxes in the TeV regime were obtained even after 14.2 hours of observation~\citep{2012Aliu}. 
At high energies the spectral energy distribution (SED) 
is severely affected by the diffuse extra--galactic background
light and the appropriate corrections to recover the intrinsic
spectral shape must be applied~\citep[e.g.][]{2008Franceschini,2011Dominguez},
before  the SED can be compared with theoretical emission
models~\citep[e.g.][]{1993Dermer,2013Bottcher}. 
In all these studies a precise measurement of the redshift is required. 
It is generally assumed that BL Lac nuclei are hosted by luminous elliptical
galaxies embedded in small groups of galaxies~\citep[e.g.][]{1990Falomo,2000Urry}.
The presence of a close companion is usually presented as 
the cause of the activation of the central engine~\citep[e.g.][]{2008Hopkins} 
however, {\it Hubble Space Telescope} observations of a large sample of BL Lac 
objects have shown than only around 50\% of them have a nearby companion
finding also truly isolated BL Lac~\citep{2000Urry}. 

Most of the previous studies on BL Lac  environment have been performed using photometric
redshifts and only in a few cases~\citep[e.g.][]{2008Lietzen,2016Farina}
the environment have been characterized with spectroscopic redshifts.
In this paper we report on the distance and environment of \TS obtained by making use of
spectroscopic redshifts gathered with the  Gran Telescopio Canarias (GTC)~\footnote{Gran Telescopio Canarias is a 
Spanish initiative with the participation of Mexico and the US University of Florida, and is in-
stalled at the Roque de los Muchachos in the island of La Palma.}. 

Throughout this paper,
we assume a flat cold dark matter cosmology with $\Omega_{\rm M}$ =0.3
and H$_0$=70 km s$^{-1}$ Mpc$^{-1}$.

\section{Observations and Data Reduction}
\label{ObandRed}
The observations were performed using the OSIRIS Multi Object 
Spectrograph (MOS) installed in the Nasmyth-B focus of the 10.4-m GTC
under the program GTC5-15BMEX (PI DRG).
The observations were carried out 
in service mode, using the 
R1000R grism. The spectrum is centred at 7430~\AA\, covering the range from 
5100 to 10000 \AA\, at a resolution of 2.62~\AA/pixel which 
translates to an effective resolution measured on strong sky lines of 10.86~\AA.
Targets are located at different positions along the dispersion axis,
changing the actual wavelength coverage;  
the common wavelength range covered by all spectra is 5400--9500~\AA.

The total observing time was divided in two observing blocks (OB) that were 
observed on February 6$^{\rm th}$ 2016.
Each OB consisted of 3 exposures of 1310 seconds 
on target 
to facilitate later removal of cosmic ray hits. 
The two OBs were accompanied by a common set of ancillary files that
included observations of G191--B2B as a standard star, bias, flat-field and arc lamps. 
Both OBs were observed with air masses lower than 1.12 under $spectroscopic$
cloud coverage and a seeing of 0.9\arcsec.

The data reduction was carried out by using a new MOS pipeline 
described in~\citet{2016arXiv160503109G}.
In short, the code reduces every MOS slit by applying the usual {\small IRAF} scripts
for long slit spectra.
To reduce the data, the three different target images were corrected by bias independently.  
Then we stacked them to obtain a single 
spectral image where the cosmic rays were successfully removed.
After that, every slitlet spectrum was calibrated in wavelength 
by using as a reference a He+Ne+Ar arc image. 
The dispersion solution is obtained for every single slitlet 
and in all the cases  rms errors lower than 0.4~\AA\, were found. 
This uncertainty produces a systematic error in the redshift calculations of
$\sim 10^{-4}$ at a wavelength of 5000\AA.  

Finally the image is calibrated in flux by using a sensitivity curve
obtained by the observation of  G191--B2B.
The output of the pipeline is a 2-D spectral image calibrated both in 
wavelength and flux.
The final signal to noise ratio around 7000\AA\, for a point source of $r\sim$21 mag. (e.g. MOS-35), 
was of around 28 in agreement with the values given by the GTC exposure time
calculator for a single OB.

After analyzing independently the two OBs we noticed that 
-- probably due to cloud coverage --  
one of them had about 40\% less counts per second than the other. 
The poor signal to noise of this latter dataset 
did not allow us to search for weak absorption lines and 
we do not use it in the following analysis.

MOS allows us to obtain the spectra of several objects within an  
effective field of view (FOV) of around 7$^{\prime}\times2^{\prime}$. 
Excluding the stars used as astrometric guides,  
within the MOS field of view we found 71 targets brighter than $r\sim$22.
We selected our targets based on the SDSS images centred   at the position of
\ts. Fifty three of the selected targets are classified as galaxies by 
SDSS, 4 of them are unclassified (including \ts), and the remaining 14
are classified as stars. 

\begin{figure}
	\includegraphics[width=\columnwidth]{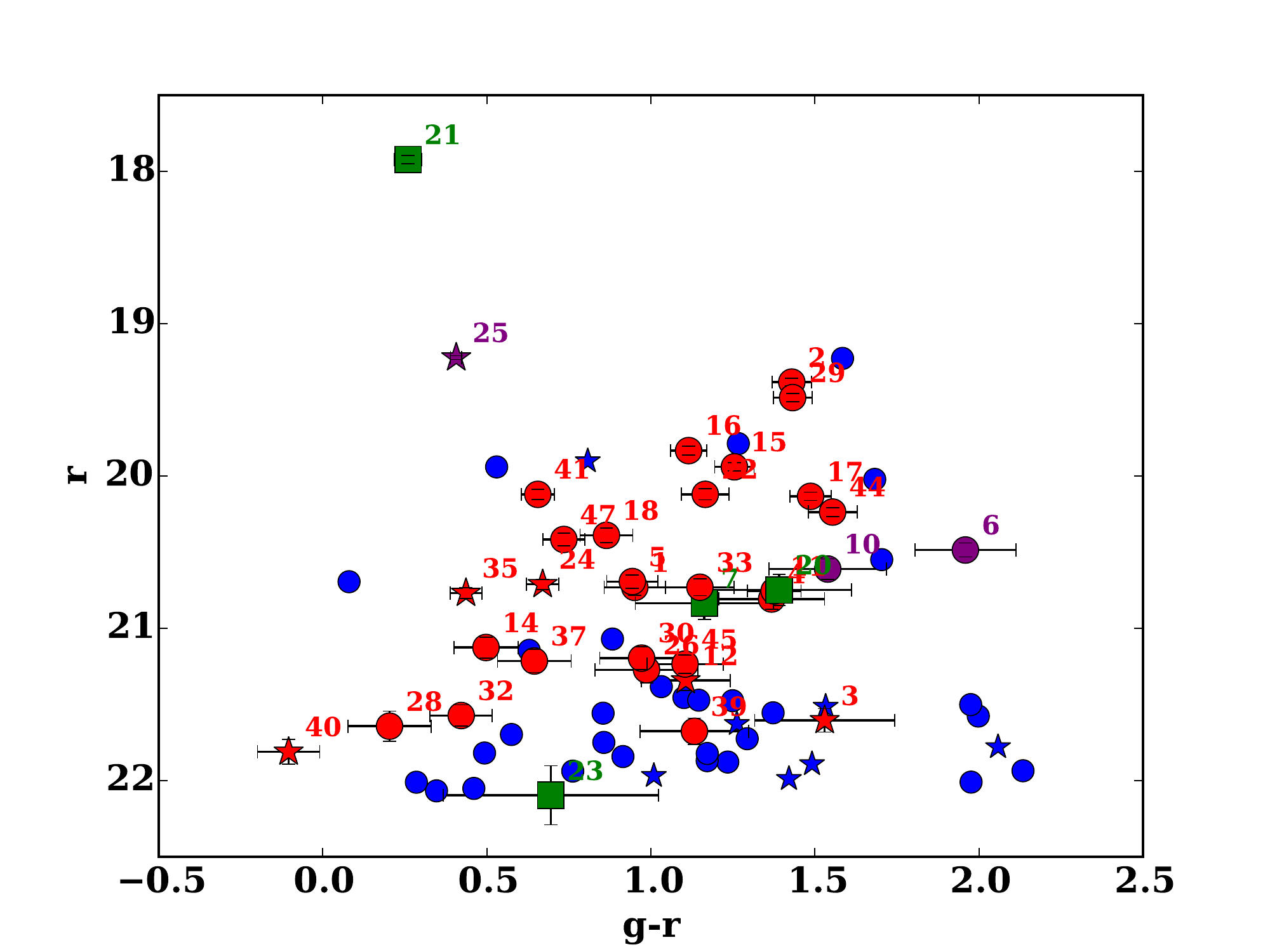}
    \caption{
Colour magnitude diagram for all objects brighter than $r\sim$22  
within the MOS field of view. 
The brightest object corresponds to \TS (MOS-21)
and it is located at the top left corner of the diagram. 
Objects classified as stars are  represented by star symbols, galaxies  by
circles and unidentified objects by squares.
In blue are the objects in the field that we did not observe.
In red, green, and purple are the targets observed by us with the slit number close to 
the corresponding symbol.
The red symbols are for objects classified as galaxies or stars, 
green symbols for objects not classified by SDSS, 
which include \TS and its nearby companion (MOS-20) and 
purple symbols are used for targets with SDSS spectra. 
The QSO at redshift 0.926 (MOS-25) was  classified  photometrically as a star 
by the SDSS photometric code.
}
    \label{fig:ColorMag}
\end{figure}

Given the nature of our program, where we do not 
know the redshift of the source, the host galaxy morphological type, or 
other extra information about the surroundings, 
we tried to cover as many objects as possible in the OSIRIS 
field of view 
taking care of the physical limitations of the 
mask making procedure, giving preference to those objects classified as galaxies.
Notice that some objects classified as stars could be distant galaxies, 
and in fact objects classified as
stars by SDSS, covered by 
slits 25 and 35, turned out  to be galaxies at $z=0.927$ and  $z=0.403$ respectively (see Table~\ref{TheTable}).
The colour magnitude diagram (CMD, Fig.~\ref{fig:ColorMag}) shows the location 
of the SDSS targets where we differentiate stars from galaxies and we marked  
those objects selected for spectroscopy.  
We cover most of the parameter space in the CMD where for 
SDSS galaxies brighter than $r$=21.5 we obtain spectroscopic redshift for 
$\sim$60\% of them. 

\begin{figure*}
	\includegraphics[width=\textwidth]{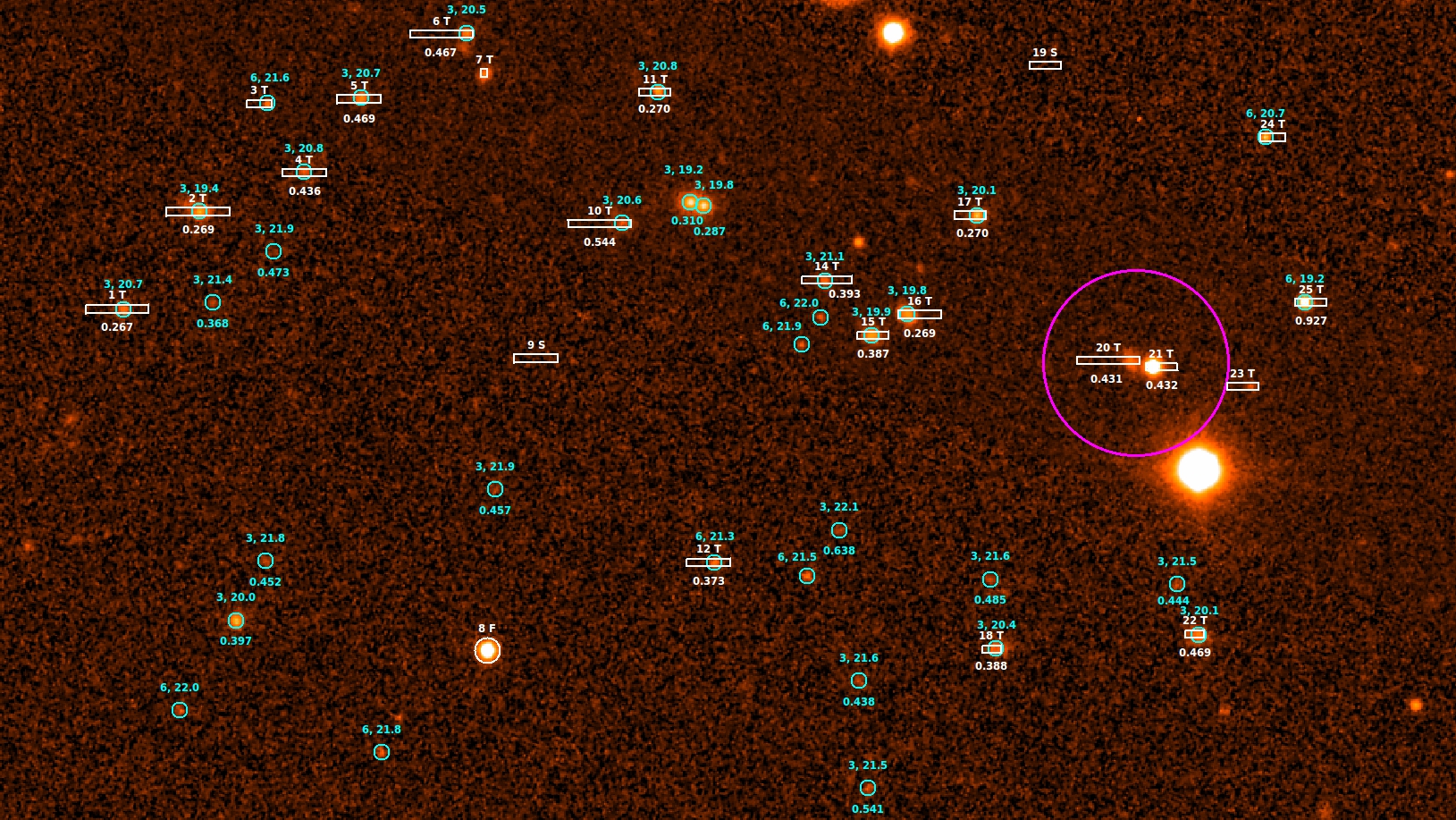}
    \caption{$r$-band SDSS image of a FOV of
      3.9\arcmin$\times$2.2\arcmin\, showing the position of the MOS
      slits (white rectangles). 
      Above the slits are the  slit numbers  
      and  the spectroscopic redshifts are indicated below them  (see Table~\ref{TheTable}).
Slits 21 and 20  mark the position of \TS and its 
nearby companion respectively, and are enclosed by a purple circle of 15\arcsec\, in radius.
      Slits 9 and 19 were used for sky spectra, and slit 8 was on top of a star 
      used as astrometric guide.
Cyan circles show the positions of SDSS objects;  
SDSS  classification (3 for galaxies, 6 for stars) 
and the $r$ magnitude are shown above each one. Below we add the photometric redshift if any. 
We only show those targets with  $18 < r < 22$.
North is at the top, East to the left, and  the FOV  covers only one of the  OSIRIS CCDs. 
}
    \label{fig:SlitPos}
\end{figure*}

Figure~\ref{fig:SlitPos} shows partially 
the slit positions on top of an \hbox{$r$-band} SDSS image, where we also marked  
the fiducial stars used for astrometry and areas free of objects used for obtaining reliable sky spectra. 
The length  of the slits goes from 1\arcsec\, to 10\arcsec\, and the width is fixed at 1.2\arcsec.   
The SDSS image shows that 
\TS -- in slit 21 -- has a nearby companion (3.8\arcsec\,  apart, Fig.~\ref{fig:SlitPos}) with unknown redshift, 
we placed one of our slits (MOS-20) 
on top of this source to find if it is physically associated to \ts.

The individual spectrum for each object was extracted from the calibrated
 2-D spectral image by using the {\small IRAF} task $apall$. 
In general the extraction window is centred on the peak 
of the continuum, however for the case of the extraction 
of the spectrum corresponding to \TS (MOS-21) and 
keeping in mind that the emission coming from the 
centre is featureless,  we extracted the 
spectrum avoiding the central pixels.
After trying with different apertures 
we end with a final extraction window centered 3 pixels (0.75\arcsec) 
to the west of the continuum peak and an aperture 
of 4 pixels (1\arcsec).
Based on the spatial profile of the emission around 5000~\AA\, we calculate that
the central pixel contribute around ~40\% of the light inside the aperture.  
Therefore by excluding that pixel from the extraction window, we improve significantly the 
observed equivalent widths.

In most of the cases the continuum is 
well detected and a 4$^{\rm th}$ order polynomial function was good enough to fit 
the trace along the dispersion axes. 
In the cases where the continuum was not well detected 
we used  the trace solutions found for a nearby object as reference. 

Once the 1-D individual spectra were extracted, we located by eye different 
spectral features. We looked for the most common lines observed in extra--galactic sources, 
both in absorption (e.g. Ca~\textsc{II} K\&H, Mg band, NaD) and in emission (e.g. Hydrogen 
recombination lines, [OII]$\lambda$3727, [OIII]$\lambda$5007, [NII]$\lambda$6583).
Once the lines were identified in a given spectrum we fit a Gaussian profile 
which returns the position of the peaks, 
 the integrated flux of the different lines and the equivalent widths 
(we used the {\small IRAF}/$splot$ command stroke $k$). 
The peaks position  was used to estimate the 
corresponding redshifts for each individual line. The final redshift
is the median of 
all measured  values. The error is given 
by the standard deviation where we have more than two
lines in the spectrum, or the difference 
between the two redshift values  
when we have only  two lines identified in the spectrum. 
The systematic error in redshift of $10^{-4}$\, 
due to uncertainties in the 
wavelength calibration described in section~\ref{ObandRed} 
was added in quadrature. 
The spectroscopic redshifts and the errors are included in Table~\ref{TheTable}.
We also included the number of lines used to estimate the redshifts and 
if they were found in emission or in absorption.

\begin{figure*}
\includegraphics[width=0.75\textwidth]{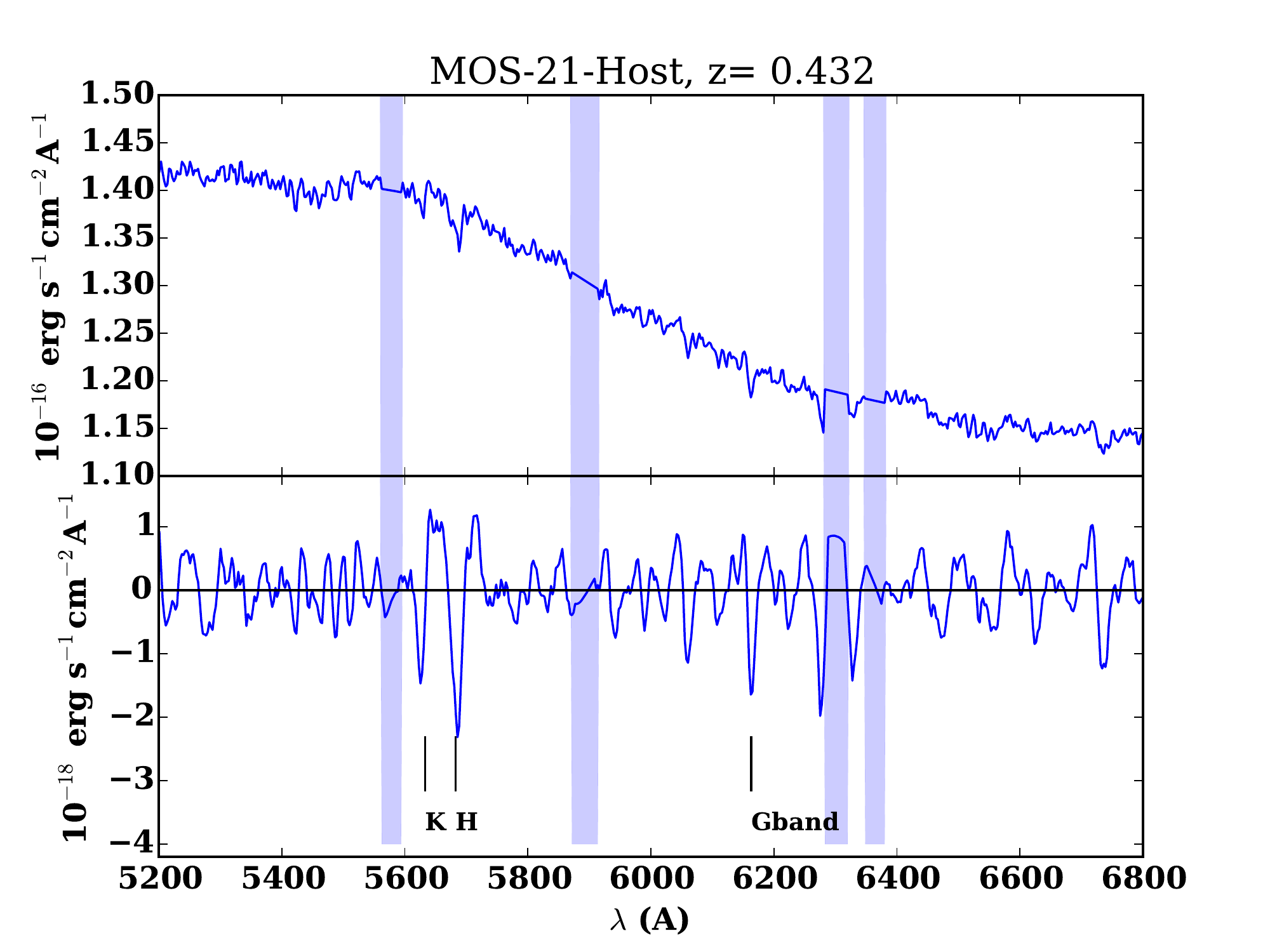}
    \caption{The observed spectrum of \TS host galaxy is presented in the top panel, 
and the continuum subtracted spectrum together with the spectral features used  
to obtain the redshift are in the bottom one. 
The spectrum of the bottom panel was smoothed to  see better the weak 
absorption lines. The areas of the spectrum where
both atmospheric lines are too strong and the telluric absorption lines 
are located, are marked with blue rectangles in the bottom panel.
Only a section of the full OSIRIS spectral coverage is plotted.  
}
    \label{fig:HostSpec}
\end{figure*}

\begin{figure}
	\includegraphics[width=\columnwidth]{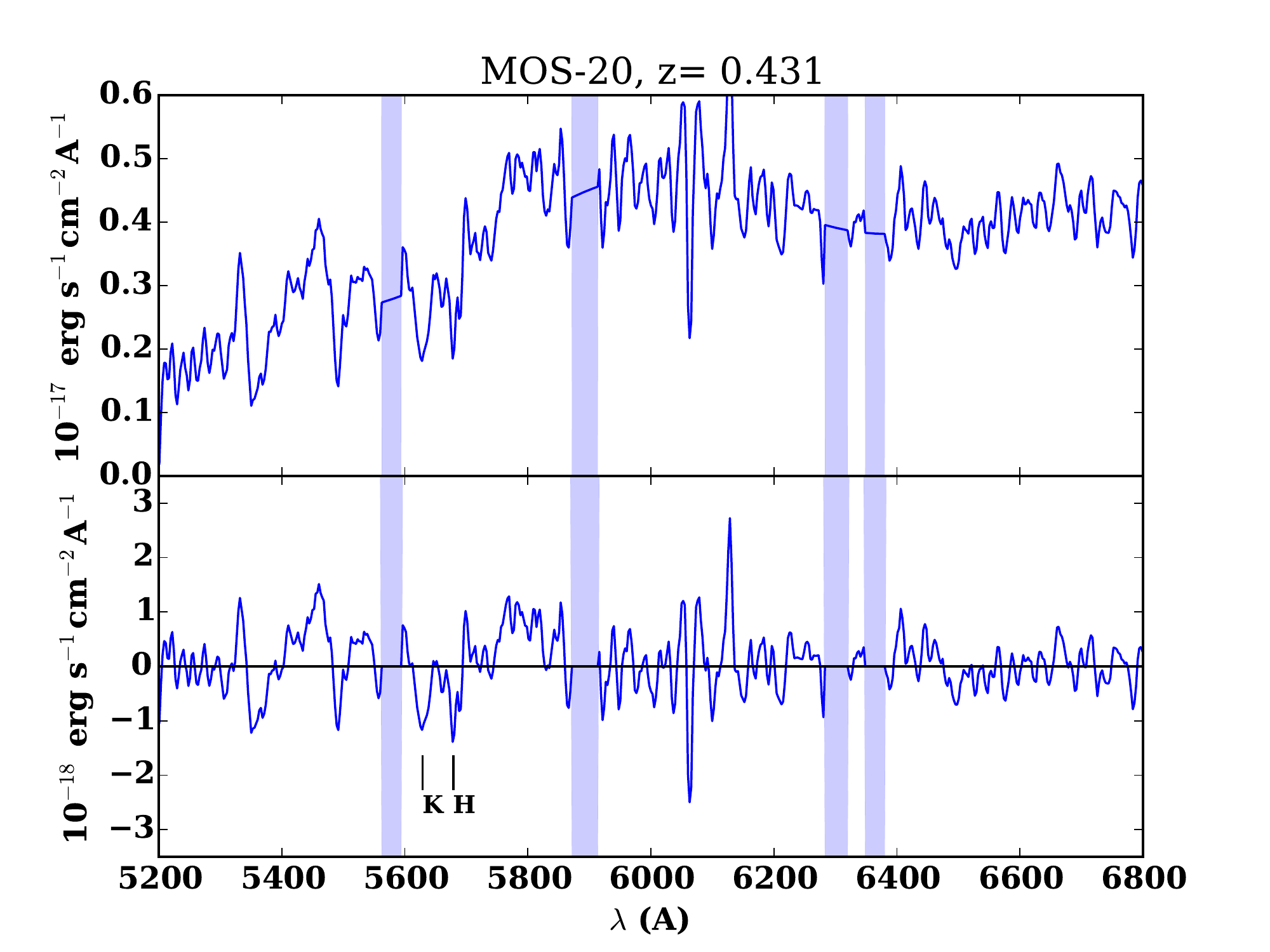}
    \caption{The spectrum of the neighbouring galaxy separated from \TS by 3.8\arcsec.
As in Figure~\ref{fig:HostSpec} we marked in the bottom panel the lines used to
estimate the redshift and the areas where strong sky lines are present.
}
    \label{fig:MOS-20}
\end{figure}

\section{Results}
\label{Results} 

\subsection{Redshift of \TS}
We obtain spectroscopic redshifts for 26 of the 35 observed targets. 
Targets for which we cannot extract the redshift include 
objects close to the slit borders and those where 
the signal to noise ratio was too low to obtain any spectral feature.  
Table~\ref{TheTable} lists the SDSS properties of the selected targets,  
the spectroscopic redshifts, and distance to \ts.
The flux calibrated spectrum of the \TS host 
galaxy is plotted in  Figure~\ref{fig:HostSpec}  (top panel).
Around 5700~\AA, 
the spectrum has a root mean square noise of \hbox{$\sim 1.1\times10^{-18}$\ergscmA}
(signal to noise ratio of \mbox{$\sim150$})
which is a factor of 9 better than that in~\cite{2012Aliu} and \cite{2013Shaw}. 
In order to see clearly the weak absorption features, 
we fit a 6$^{\rm th}$ order polynomial curve to the original spectra plus a 
Savitzky--Golay filter~\citep[e.g.][]{1989Press}\footnote{The Savitzky Golay filter is a particular type of low-pass filter.
We make use of the routine provided by the SciPy organization at SciPy.org.
Information of the algorithm and multiple references can be found in the SciPy pages.}
which removes very broad spectral features  
due to residuals produced in the sky substraction process,  
and therefore they are not related with any astrophysical source. 
The result of that process is plotted in the bottom panel of Fig.~\ref{fig:HostSpec}.  
We mark in the plot three absorption lines used to 
obtain the redshift of $z=0.432\pm0.002$. 
These spectral features from Ca~\textsc{II}~K\&H and the G--band,
have their origin in stellar atmospheres, therefore 
they are coming from the \TS host galaxy, 
and not from any intervening material in the line of sight. 
Table~\ref{LineProp} shows the line fluxes and equivalent widths obtained
  from the Gaussian profile fit. In all cases the lines were detected above 
3$\sigma$ level.

\begin{table}
	\centering
	\caption{Line fluxes and equivalent widths of the three absorption lines detected in
          the host galaxy of \ts. }
	\label{LineProp}
	\begin{tabular}{ccc} 
\hline
\ & Line Flux & Equivalent Width\\
\ & ($\times 10^{-17}$\ergscmA) & (\AA)\\ \hline
Ca II K & 3.46 $\pm$ 1.07 & 0.25 $\pm$  0.08 \\
Ca II H & 8.63 $\pm$ 1.44 & 0.62 $\pm$  0.10 \\
G band  & 3.37 $\pm$ 0.94 & 0.28 $\pm$  0.08 \\	\hline
\end{tabular}
\end{table}

The spectra of the nearby companion is shown in Fig.~\ref{fig:MOS-20}
and based on the observed absorption lines we obtain a $z=0.431\pm0.002$. 
The angular separation between these two sources is of 3.8\arcsec\, which 
corresponds to $\sim$21~kpc at the redshift of the BL Lac.

We calculate the radial velocity dispersion distribution taking  as
reference the recessional velocity of \TS (Fig.~\ref{fig:Vdist}).
The distribution includes both photometric and spectroscopic redshifts.
The spectroscopic data comes from our MOS observations and the photometric redshifts were obtained from 
the SDSS data release 12~\citep{2016Beck} which includes all the 
galaxy--type objects within the OSIRIS FOV  (7$^{\prime}\times2^{\prime}$ centered in \ts).  
We did not find any significant structure around \ts.
The difference between the recession velocity of \TS and its nearby companion is
$\sim$300 \kms\, which is similar to 
the dispersion velocities observed in galaxy groups. 

\subsection{Redshift of Galaxies in the OSIRIS FOV}
\begin{figure}
	\includegraphics[width=\columnwidth]{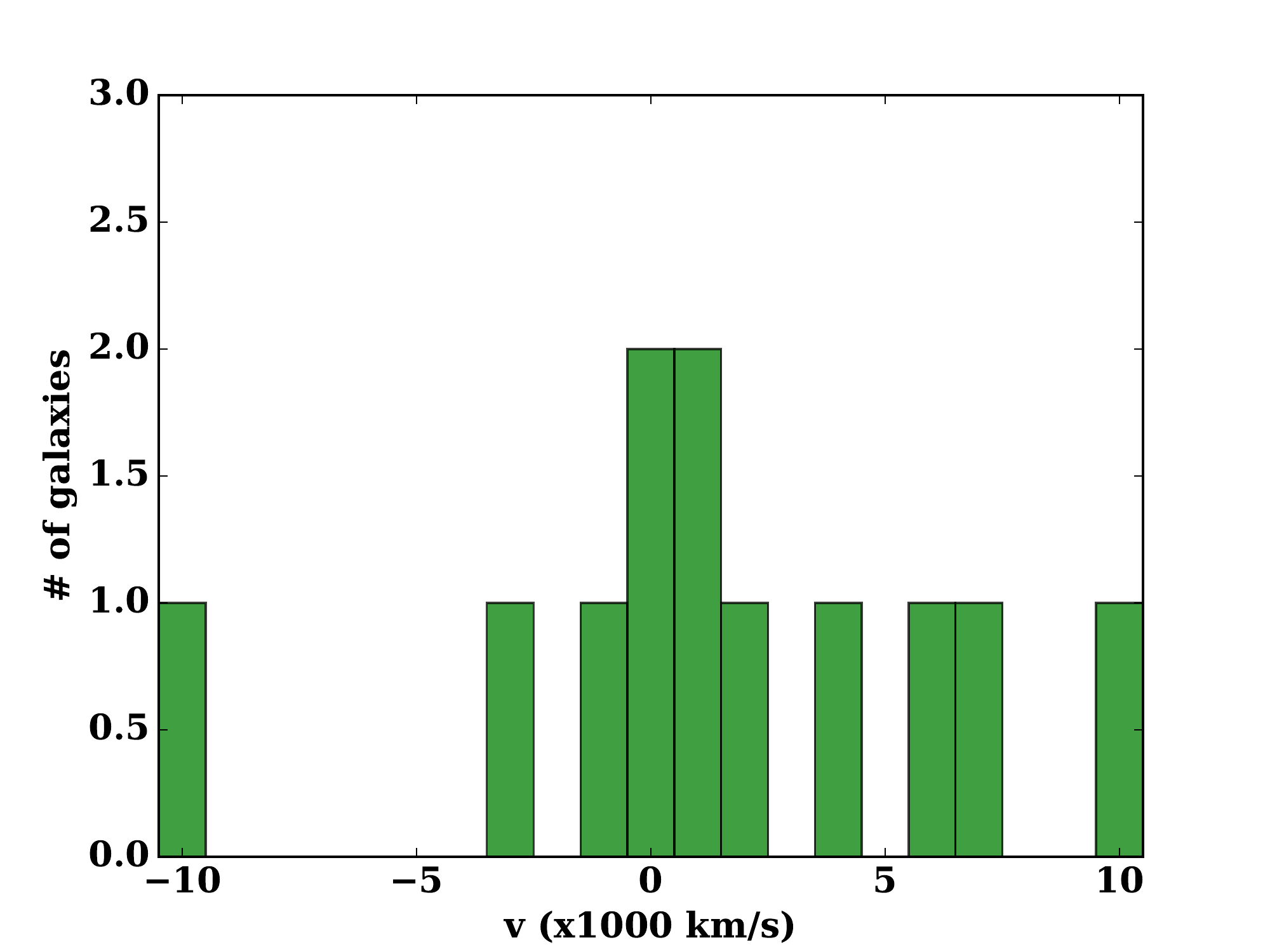}
    \caption{
Radial velocity distribution taking as reference the 
recessional velocity of \ts. 
The bin size is of  1000 km~s$^{-1}$  and we included both spectroscopic and photometric
redshifts. 
Only \TS and its close companion lie inside the central bin.
}
    \label{fig:Vdist}
\end{figure}

\begin{figure}
	\includegraphics[width=\columnwidth]{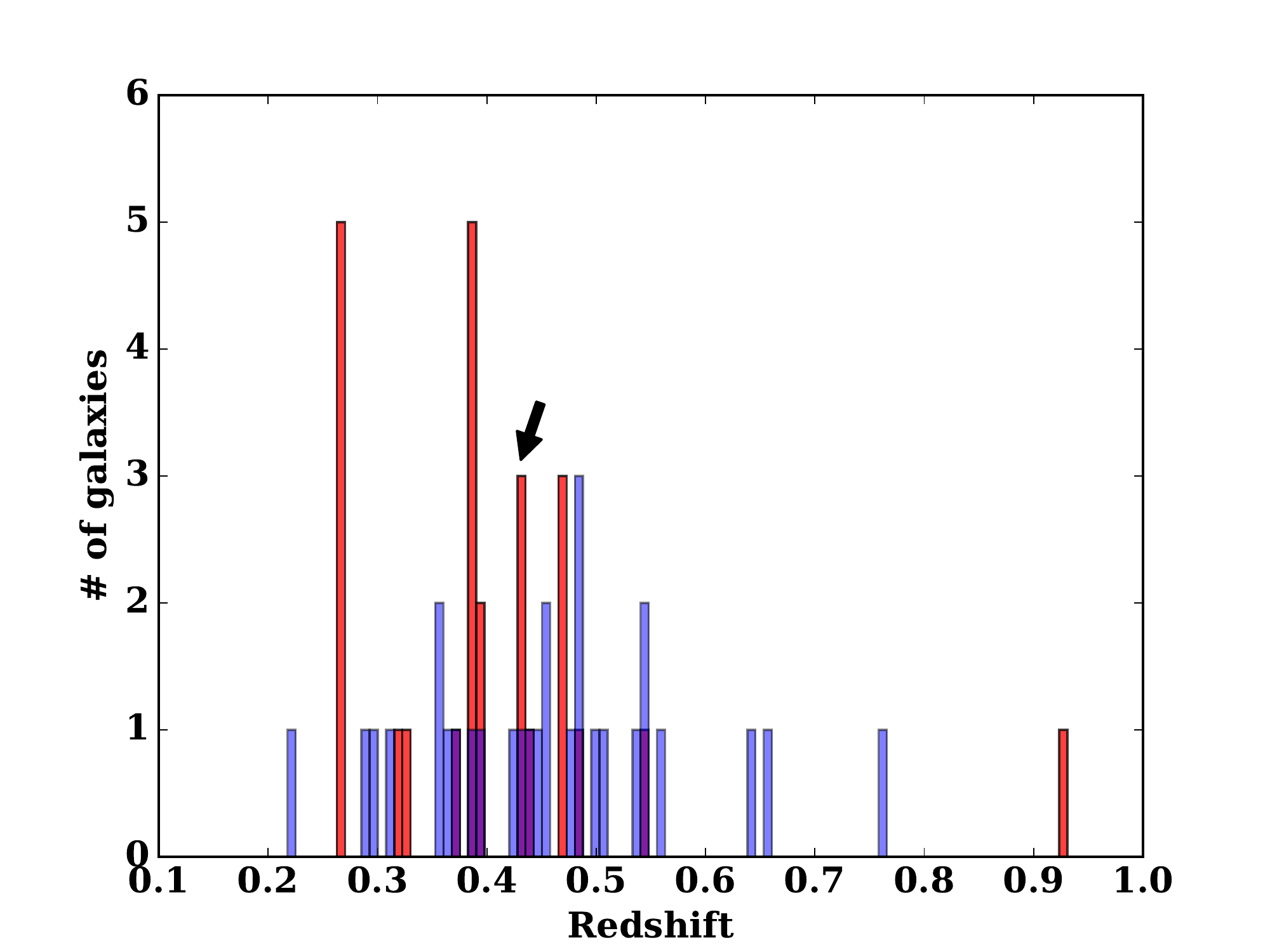}
    \caption{Spectroscopic  (red) 
and photometric (blue) redshift  distributions 
of the targets within the OSIRIS FOV.
The bin size is  0.0075 and the bin containing \TS and its 
close companion is marked with an arrow. 
Two small groups of galaxies at redshifts 0.28 and 0.39 
are identified. 
}
    \label{fig:Zdist}
\end{figure}

To study the existence of galaxy groups within the OSIRIS FOV  
we create a histogram showing 
the redshift distribution based on photometric and 
spectroscopic redshifts (Fig.~\ref{fig:Zdist}).
As shown in Figure~\ref{fig:Vdist}, 
we did not find any significant over density around \ts, either using the 
spectroscopic or photometric redshifts. 
Within the redshift bin where \TS is located, 
we find just another galaxy (MOS-44) with similar redshift $0.4347\pm0.0006$ 
but separated by $\sim$270\arcsec. This angular separation corresponds to a 
projected physical distance of 1.5~Mpc, therefore this galaxy is 
too far away to be related to the nuclear activity of \ts. 

Two small peaks appear in the spectroscopic distributions at 
0.28 and 0.39 which are not related with \ts. 
The minimum and maximum distances between the members belonging 
to the group at $z$=0.28 are 86, and 585 kpc respectively.
The corresponding values for the group at $z$=0.39 are 286 kpc, and 1.7 Mpc.
The spectroscopic study of these two groups will be presented in a
forthcoming paper.

\section{Conclusions}

By carefully extracting the GTC spectra  of \TS 
avoiding the featureless central engine 
we assign a redshift of $z$=0.432$\pm$0.002 
to its host galaxy.
We find a neighbouring galaxy just 21~kpc apart with similar redshift
that could be the cause of the BL Lac  nuclear activity observed in \ts. 
In fact, the presence of nearby companions 
separated by tens of kpc have been observed in other BL Lac objects
pointing to the possible connection between the 
gas fueling towards the nuclei due to a close encounter 
and the consequent triggering the nuclear activity 
~\citep[e.g.][]{1996Falomo,2016Farina}.
 
Our results show that in a FOV of $7\arcmin\times2\arcmin$ exist two groups of
galaxies with mean redshifts which are quite different from the spectroscopic redshift of
the BL Lac. This result indicates that conclusions obtained by 
associating the redshift of a given object based on 
the presence of a galaxy group within the studied FOV  
\cite[e.g.][]{2015Muriel} must be used with caution.

The fact that we now have a precise spectroscopic redshift determination for \TS and taking into account that
photons from this object have been detected above 10~GeV, makes
\TS  an interesting target for Cherenkov telescopes. 
                                           
\begin{table*}
	\centering
	\caption{Targets observed with  GTC-MOS. Columns   1 to 6 give: 
slit number (ID), SDSS coordinates,  type (0 unclassified, 3 for galaxies, 6
for stars), SDSS $r$ band, and SDSS $g-r$ colour. 
The  spectroscopic redshift ($z$), its error and the number of lines (N) used for the
redshift determination are given in columns 7 and 8. 
Column 8 also indicates whether the lines were observed in absorption (A) or in
emission (E). The angular separation ($\theta$) to \TS is in column 9. For reference, at the
redshift of \ts, 1\arcsec corresponds to 5.625 kpc.
The slits on  fiducial stars or those used for sky measurements are not shown.
}
	\label{TheTable}
	\begin{tabular}{lcccccccrc} 
		\hline
ID & RA($^{\rm o}$) & DEC($^{\rm o}$)                 & Type & $r$\,(mag.) & $g-r$\,(mag.) & $z$\  \  \ \ \ & N & $\theta$ (\arcsec) &  Comments \\    
\hline
MOS-1 &  137.30323 & 23.18951 &  3 &  20.73$\pm$0.05 & 0.95$\pm$0.09 & 0.2670$\pm$0.0002 &  4/E & 182.4 & --\\
MOS-2 &  137.29928 & 23.19389 &  3 &  19.38$\pm$0.02 & 1.43$\pm$0.06 & 0.269$\pm$0.0010 &  6/A & 169.0 & --\\
MOS-3 &  137.29633 & 23.19840 &  6 &  21.60$\pm$0.08 & 1.53$\pm$0.21 & -- &  -- & 160.3 & (d)\\
MOS-4 &  137.29408 & 23.19562 &  3 &  20.81$\pm$0.07 & 1.37$\pm$0.16 & 0.436$\pm$0.0060 &  2/A & 151.1 & --\\
MOS-5 &  137.29140 & 23.19894 &  3 &  20.69$\pm$0.04 & 0.94$\pm$0.08 & 0.4685$\pm$0.0003 &  5/E & 143.2 & --\\
MOS-6 &  137.28744 & 23.20185 &  3 &  20.49$\pm$0.05 & 1.96$\pm$0.15 & 0.4665$\pm$0.0003 &  3/A & 131.5 & (SDSS)\\
MOS-7 &  137.28540 & 23.19999 &  0 &  20.84$\pm$0.11 & 1.16$\pm$0.21 & -- &  -- & 123.0 & (e)\\
MOS-10 &  137.27964 & 23.19333 &  3 &  20.61$\pm$0.07 & 1.54$\pm$0.18 & 0.544$\pm$0.0030 &  4/A & 98.8 & (SDSS)\\
MOS-11 &  137.27697 & 23.19924 &  3 &  20.75$\pm$0.04 & 1.38$\pm$0.08 & 0.27$\pm$0.0080 &  5/A & 93.6 & --\\
MOS-12 &  137.27430 & 23.17811 &  6 &  21.34$\pm$0.07 & 1.11$\pm$0.14 & 0.373$\pm$0.0030 &  2/E & 81.5 & --\\
MOS-14 &  137.26851 & 23.19081 &  3 &  21.13$\pm$0.07 & 0.50$\pm$0.10 & 0.3925$\pm$0.0002 &  5/E & 58.2 & --\\
MOS-15 &  137.26625 & 23.18832 &  3 &  19.94$\pm$0.03 & 1.25$\pm$0.06 & 0.3868$\pm$0.0001 &  4/E & 49.3 & --\\
MOS-16 &  137.26397 & 23.18925 &  3 &  19.83$\pm$0.03 & 1.12$\pm$0.05 & 0.269$\pm$0.0070 &  4/A & 41.4 & --\\
MOS-17 &  137.26150 & 23.19371 &  3 &  20.13$\pm$0.03 & 1.49$\pm$0.06 & 0.2698$\pm$0.0004 &  6/A & 36.7 & --\\
MOS-18 &  137.26049 & 23.17430 &  3 &  20.39$\pm$0.05 & 0.86$\pm$0.08 & 0.388$\pm$0.0002 &  4/E & 43.9 & --\\
MOS-20 &  137.25487 & 23.18718 &  0 &  20.75$\pm$0.10 & 1.39$\pm$0.22 & 0.431$\pm$0.0020 &  2/A & 3.8 & (b)\\
MOS-21 &  137.25216 & 23.18690 &  0 &  17.92$\pm$0.03 & 0.26$\pm$0.04 & 0.432$\pm$0.0020 &  3/A & 0.0 & (a)\\
MOS-22 &  137.25051 & 23.17490 &  3 &  20.12$\pm$0.04 & 1.17$\pm$0.07 & 0.4692$\pm$0.0004 &  5/EA & 32.7 & --\\
MOS-23 &  137.24843 & 23.18602 &  0 &  22.10$\pm$0.19 & 0.70$\pm$0.33 & -- &  -- & 15.2 & (c)\\
MOS-24 &  137.24657 & 23.19721 &  6 &  20.71$\pm$0.03 & 0.67$\pm$0.05 & -- &  -- & 34.8 & (c)\\
MOS-25 &  137.24483 & 23.18979 &  6 &  19.22$\pm$0.01 & 0.41$\pm$0.02 & 0.927$\pm$0.0030 &  2/E & 29.0 & (SDSS)\\
MOS-26 &  137.23555 & 23.16875 &  3 &  21.27$\pm$0.09 & 0.99$\pm$0.16 & 0.383$\pm$0.0070 &  3/E & 78.0 & --\\
MOS-28 &  137.23118 & 23.17396 &  3 &  21.64$\pm$0.10 & 0.20$\pm$0.13 & -- &  -- & 84.4 & (c)\\
MOS-29 &  137.22870 & 23.18817 &  3 &  19.49$\pm$0.03 & 1.43$\pm$0.06 & 0.384$\pm$0.0010 &  4/E & 86.1 & --\\
MOS-30 &  137.22499 & 23.17738 &  3 &  21.20$\pm$0.07 & 0.97$\pm$0.13 & 0.32$\pm$0.0030 &  2/A & 102.6 & --\\
MOS-32 &  137.22047 & 23.17773 &  3 &  21.57$\pm$0.07 & 0.42$\pm$0.09 & -- &  -- & 118.2 & (c)\\
MOS-33 &  137.21864 & 23.18462 &  3 &  20.73$\pm$0.05 & 1.15$\pm$0.10 & 0.481$\pm$0.0050 &  3/A & 122.4 & --\\
MOS-35 &  137.21341 & 23.18033 &  6 &  20.77$\pm$0.03 & 0.44$\pm$0.05 & 0.064$\pm$0.0040 &  2/A & 142.2 & --\\
MOS-37 &  137.20116 & 23.17549 &  3 &  21.21$\pm$0.08 & 0.64$\pm$0.11 & 0.3929$\pm$0.0001 &  4/E & 187.6 & --\\
MOS-39 &  137.18953 & 23.19677 &  3 &  21.68$\pm$0.09 & 1.13$\pm$0.17 & -- &  -- & 228.5 & (e)\\
MOS-40 &  137.18827 & 23.18101 &  6 &  21.81$\pm$0.08 & -0.10$\pm$0.09 & -- &  -- & 232.1 & (e)\\
MOS-41 &  137.18586 & 23.17341 &  3 &  20.12$\pm$0.03 & 0.66$\pm$0.05 & -- &  -- & 242.9 & (e)\\
MOS-44 &  137.17782 & 23.18492 &  3 &  20.24$\pm$0.03 & 1.55$\pm$0.07 & 0.4347$\pm$0.0006 &  5/A & 269.3 & --\\
MOS-45 &  137.17472 & 23.16406 &  3 &  21.23$\pm$0.06 & 1.10$\pm$0.12 & 0.3856$\pm$0.0002 &  5/E & 286.8 & --\\
MOS-47 &  137.16818 & 23.18464 &  3 &  20.42$\pm$0.04 & 0.73$\pm$0.06 & 0.3230$\pm$0.0001 &  4/E & 304.0 & --\\
		\hline
	\end{tabular}
\\
The  magnitudes and colours of the SDSS unclassified objects (type=0) were 
obtained by doing photometry on the SDSS images.
(a) Main target, (b) a galaxy close to \TS, 
(c) close to the borders, impossible to get the spectra; (d) field star; 
(e) no spectral features detected,
(SDSS) we agree -- within the errors -- with the spectroscopic redshift given by SDSS.
\end{table*}

\section*{Acknowledgements}

We thank the support team at GTC, 
and an anonymous referee for suggestions that helped to improve the clarity of the paper. 

This work is partly financed 
by CONACyT (Mexico) research grants CB-2010-01-155142-G3
(PI:YDM) and CB-2011-01-167281-F3 (PI:DRG). SCL thanks
CONACyT for her studentship.

Funding for SDSS-III has been provided by the Alfred P. Sloan Foundation, 
the Participating Institutions, the National Science Foundation, 
and the U.S. Department of Energy Office of Science. The SDSS-III web site is http://www.sdss3.org/.










\bsp	
\label{lastpage}
\end{document}